\documentclass[twocolumn,aps,showpacs,pra]{revtex4}
\usepackage{epsfig}
\usepackage[english]{babel}
\usepackage{latexsym}
\usepackage{graphics}
\usepackage{subfigure}
\usepackage{epsfig}
\usepackage{graphicx}
\usepackage{dcolumn}
\usepackage{amsmath}

\begin{document}
\title{Reading molecular messages from the intersections of high-order harmonic spectra at different orientation angles}
\author{Y. J. Chen$^{1,2}$,  J. Liu$^{2,3*}$, and Bambi Hu$^{1,4}$}

\date{\today}

\begin{abstract}
We investigate  the orientation dependence of high-order harmonic
generation (HHG) from H$_2^+$ with different internuclear distances
irradiated by intense laser fields both numerically and
analytically. The calculated molecular HHG spectra are found to be
sensitive to molecular axis orientation relative to incident laser
field polarization and internuclear separation. In particular, the
spectra calculated for different orientation angles demonstrate a
kind of intersection,  which is identified as arising due to
intramolecular two-center  interference in the HHG. The striking
"intersection" phenomenon can be used to probe the molecular
instantaneous structure.
\end{abstract}
\affiliation{1. Department of Physics, Centre for Nonlinear Studies,
and The Beijing-Hong Kong-Singapore Joint Centre for Nonlinear and
Complex Systems (Hong Kong), Hong Kong Baptist University, Kowloon
Tong, Hong Kong, China\\2.Institute of Applied Physics and
Computational
Mathematics, Beijing 100088, China\\
3.Center for Applied Physics and Technology, Peking University,
Beijing 100084, China\\4. Department of Physics, University of
Houston, Houston, Texas 77204-5005.}  \maketitle

High-order harmonic generation (HHG) from aligned molecules in
strong laser fields of femtosecond duration has proven to be a
powerful tool for resolving and controlling the processes in an
ultrafast time 
scale. For instance, recent experiments showed that calibrating the
molecular recollision electronic wave packet by a reference atom,
the HHG spectra can be used to image molecular
orbital\cite{jitatani}, and measuring the interference minima in the
HHG spectra\cite{lein1,lein2,lein3,Bandrauk}, the HHG can also be
used to probe molecular instantaneous structure\cite{Tsuneto,vozzi}.
Moreover, the investigations on the molecular orbital
tomography\cite{J. Itatani,Levesque,S. Patchkovskii,R. Torres,Le}
and the effects of two-center interference in the HHG\cite{Anh-Thu
Le,Faria,Kanai,Ciappina,Wagner,Ciappina2,Usachenko} are leading
valuable  insights into the mechanism of atomic and molecular HHG.

Nevertheless, theoretical studies demonstrated that in some cases,
the molecular properties already enter the recollision electronic
wave packet. Accordingly, the spectral amplitude of the molecular
recollision electronic wave packet is largely different from its
reference atom in some energy regions\cite{cyj}. This implies that
accurately calibrating the molecular recollision electronic wave
packet can be difficult for some species of molecules. In addition,
it was revealed that both two-center interference\cite{lein1} and
the interference between different recombination electron
trajectories\cite{Lewenstein} are responsible for the suppressed
harmonics at certain orders. As a result, the minima in harmonic
spectra may shift as the laser intensity changes\cite{cyj2}.

In the present paper, numerically investigating the orientation
dependence of the HHG from H$_2^+$ with different internuclear
distances, we find that the molecular HHG spectra are sensitive to
molecular axis orientation relative to incident laser field
polarization as well as to the internuclear separation. In
particular, the spectra calculated for different orientation angles
demonstrate a kind of intersection, which is identified as arising
due to intramolecular two-center interference in the  HHG. Compared
to the interference-related minima in the HHG spectra, the
intersections of the harmonic spectra are easy to identify in
practice. This finding is advocated for promising applications as a
prospective tool to probe the molecular structure and dynamics.

Let us consider the Hamiltonian of   H$_2^+$ as (the atom units of
$\hbar=e=m_e=1$ are used throughout this paper)
\begin{equation}
H(t)=\mathbf{p}^2/2+V(\mathbf{r})-\mathbf{r}\cdot \mathbf{E}(t),
\end{equation}
where $V(\mathbf{r})$ is the Coulomb potential and $\mathbf{E}(t)$
is the external electric field.  In the 2D case, the Coulomb
potential is written as
$V(x,y)=\frac{-Z}{\sqrt{0.5+({x+R/2})^{2}+y^2}}+\frac{-Z}{\sqrt{0.5+({x-R/2})^{2}+y^2}}$,
where $Z$ is the effective charge, and $R$ is the internuclear
separation, (for $Z=1$ and  $R=2$ a.u., the ground state energy for
H$_2^+$ reproduced here is $I_p=1.11$ a.u.).  In this paper, we
assume that the molecular axis is coincident with the x-axis and the
external field $\mathbf{E}(t)=E\vec{e}\sin(\omega_{0} t)$ is
linearly polarized with an orientation angle of $\theta$ to the
molecular axis. $E$ and $\omega_{0}$ are the amplitude and angular
frequency of the laser field. $\vec{e}$ is the unit vector along the
laser polarization.  Our calculation will be considered for $780$ nm
trapezoidally shaped laser pulses with a total duration of $10$
optical cycles and linear ramps of three optical cycles.
Numerically, the  Schr\"odinger equation with the above Hamiltonian
$H(t)$¡¡\ is solved by the operator-splitting method with 2048 time
steps per optical cycle.  The numerical convergency is checked using
a finer grid. The coherent part of the harmonic spectrum is obtained
from the Fourier transformed dipole acceleration expectation value,
and only the harmonics polarized parallel to the incoming field are
considered\cite{lein1}.

\begin{figure}[t]
\begin{center}
\rotatebox{0}{\resizebox *{8.5cm}{8.cm} {\includegraphics
{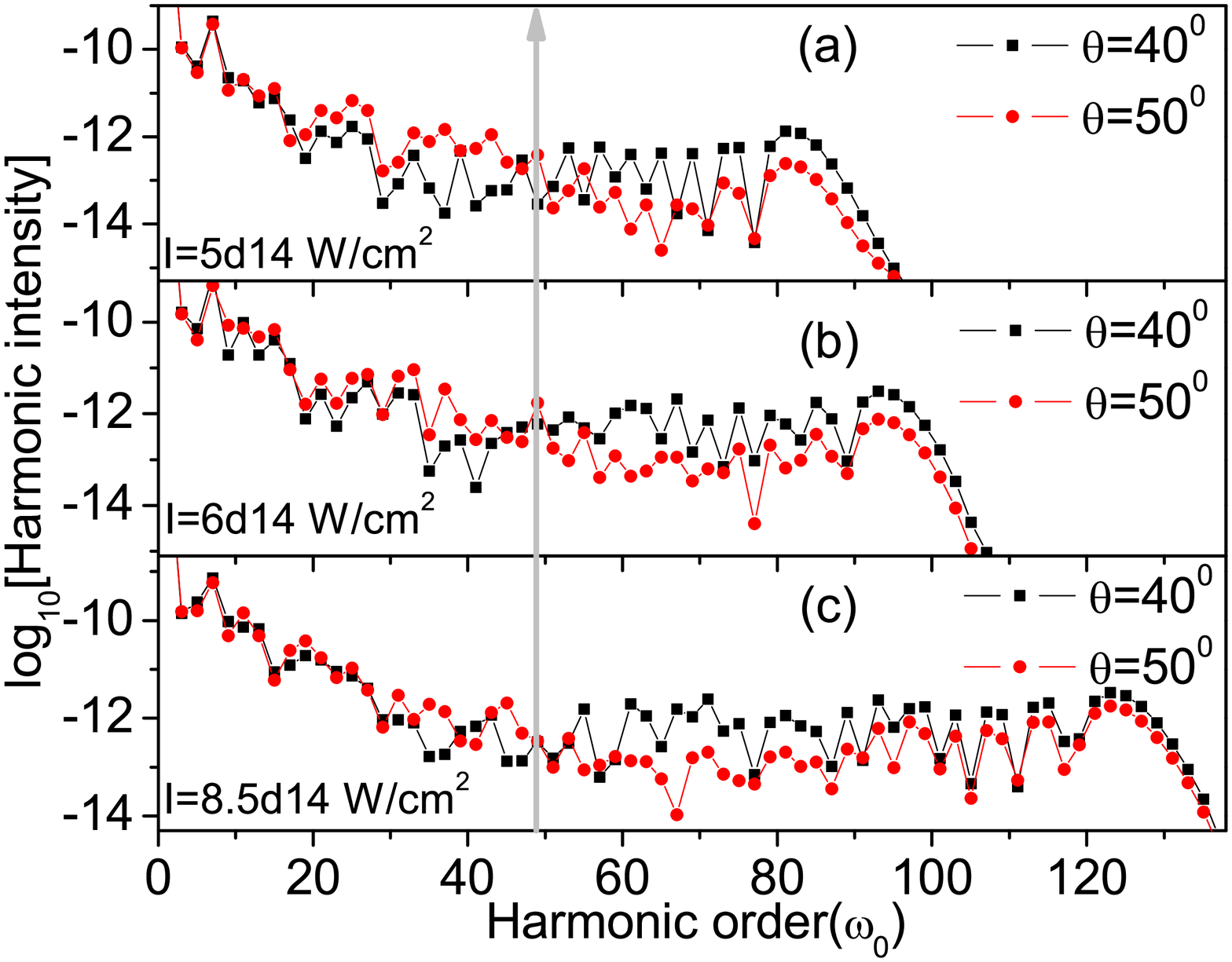}}}
\end{center}
\caption{Harmonic spectra  of 2D H$_2^+$  with $Z=1$, $R=2$ a.u. at
different laser intensities and orientation angles $\theta$,
obtained by the exact numerical calculation. } \label{fig.1}
\end{figure}
To analytically investigate the molecular HHG, we use  the
Lewenstein model\cite{Lewenstein} and consider the acceleration
effect of the bound potential\cite{jitatani,lein1}. The
time-dependent dipole moment is given by\cite{Kanai,Ciappina}
\begin{eqnarray}
\mathbf{x}(t)=&
i\int^\infty_{0}d\tau\big(\frac{\pi}{\epsilon+i\tau/2}\big)
\mathbf{d}^{*}_{rec}(\mathbf{p}_{st}-\mathbf{A}(t))e^{-iS(p_{st},t,\tau)}\nonumber\\
&\times
\mathbf{E}(t-\tau)\cdot\mathbf{d}_{ion}(\mathbf{p}_{st}-\mathbf{A}(t-\tau))+c.c.,
\end{eqnarray}
where
$S(p_{st},t,\tau)=\int^t_{t-\tau}dt''\big[\big(\mathbf{p}_{st}-\mathbf{A}(t'')\big)^{2}/2+I_p\big]$
is the semiclassical action, $\mathbf{A}(t)=-\int\mathbf{E}(t')dt'$
is the vector potential of the external field, and
$\mathbf{p}_{st}=\int^t_{t-\tau}dt''\mathbf{A}(t'')/\tau$ is the
canonical momentum corresponding to the stationary value.
$\mathbf{d}_{ion}(\mathbf{p})=\langle\mathbf{p}|\mathbf{\hat{r}}|0\rangle=(2\pi)^{-3/2}\int^\infty_{-\infty}
d\mathbf{r}exp(-\mathbf{p}\cdot\mathbf{r})\mathbf{r}\langle
\mathbf{r}|0\rangle$ is the bound-free dipole transition matrix
element between the molecular ground state $|0\rangle$ and the
continuum $|\mathbf{p}\rangle$ in the ionization step, and
$\mathbf{d}_{rec}(\mathbf{p})=\langle\mathbf{p_{k}}|\mathbf{\hat{r}}|0\rangle=(2\pi)^{-3/2}\int^\infty_{-\infty}
d\mathbf{r}exp(-\mathbf{p_{k}}\cdot\mathbf{r})\mathbf{r}\langle
\mathbf{r}|0\rangle$ is that in the recombination step which
considers the effect of the electron acceleration in the vicinity of
the parent ion before recombination, i.e., the effective momentum
$\mathbf{p_k}=\sqrt{\mathbf{p}^{2}+2I_p}\mathbf{p}/|\mathbf{p}|$ is
adopted to describe the acceleration
effect\cite{jitatani,lein1,Kanai,Levesque}.

The wave function of the valence orbital of the H$_2^+$ molecule
with $1s\sigma_{g}$ symmetry, investigated here, is expressed in
the LCAO-MO approximation
\begin{eqnarray}
\mathbf{\psi}_{1s\sigma_{g}}(\mathbf{r})=N_{1s\sigma_{g}}[\mathbf{\phi}_{1s}(\mathbf{r}+\mathbf{R}/2)+\mathbf{\phi}_{1s}(\mathbf{r}-\mathbf{R}/2)],
\end{eqnarray}
where $N_{1s}$ is the normalization factor, $\mathbf{\phi}_{1s}$ is
the atomic $1s$ orbital in the configuration space, and $\mathbf{R}$
is the vector between the two atomic cores of the molecule. Then,
the dipole transition moment for H$_2^+$  can be written as
\begin{eqnarray}
\mathbf{d}_{1s\sigma_{g}}(\mathbf{p})=2iN'_{1s\sigma_{g}}[-\cos(\mathbf{{p}}\cdot
\mathbf{{R}}/2)\mathbf{{d}}_{1s}(\mathbf{{p}})\nonumber\\+\sin(\mathbf{{p}}\cdot
\mathbf{{R}}/2)\tilde{\mathbf{{\phi}}}_{1s}(\mathbf{p})\mathbf{{R}}/2],
\end{eqnarray}
where $\mathbf{{d}}_{1s}(\mathbf{{p}})$ is the atomic dipole
moment from the $1s$ orbital, and
$\tilde{\mathbf{{\phi}}}_{1s}(\mathbf{p})$ is the $1s$ wave
function in the momentum space.

\begin{figure}[t]
\begin{center}
\rotatebox{0}{\resizebox *{8.5cm}{8.cm} {\includegraphics
{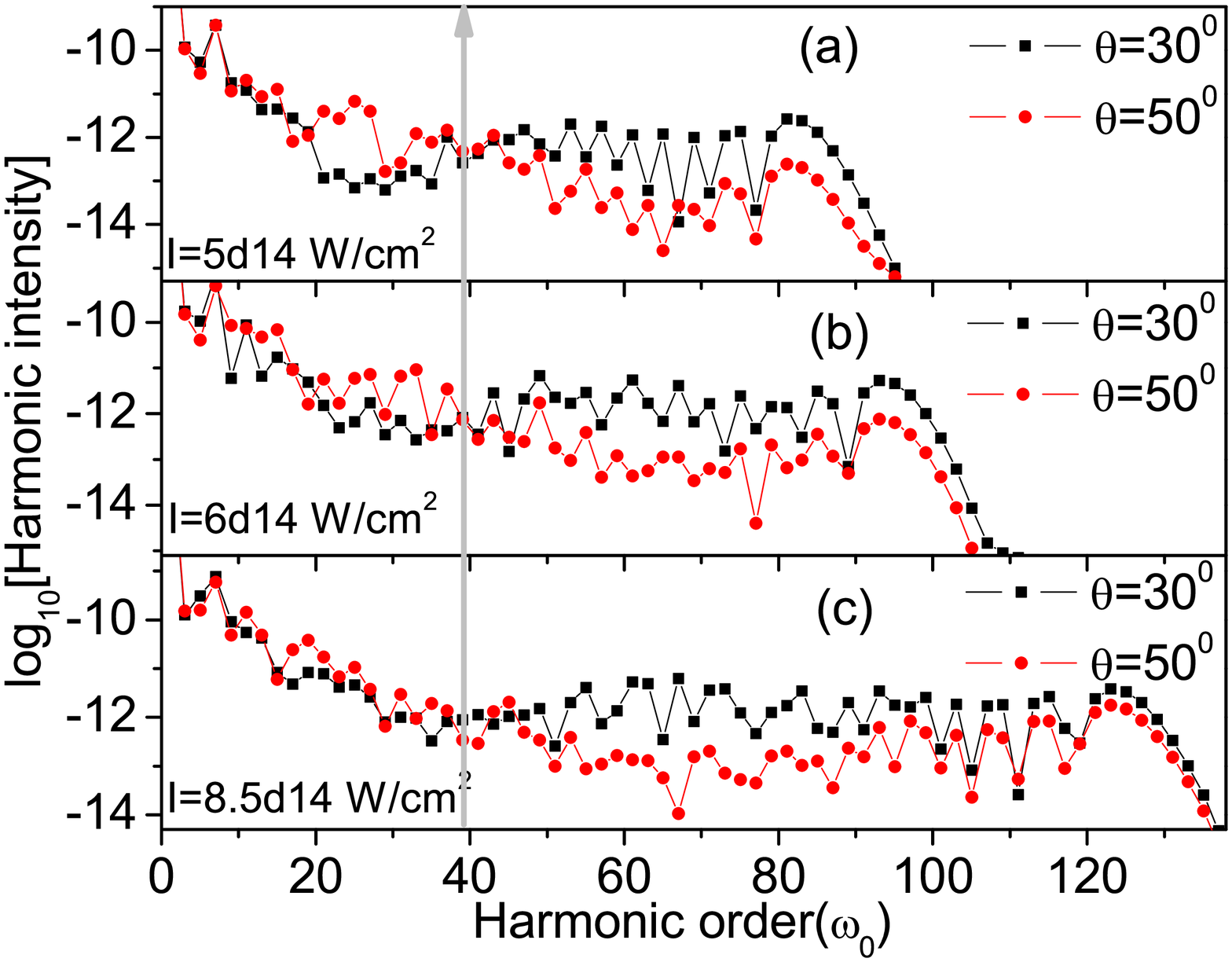}}}
\end{center}
\caption{Harmonic spectra  of 2D H$_2^+$  with $Z=1$, $R=2$ a.u. at
different laser intensities and orientation angles $\theta$,
obtained by the exact numerical calculation.  } \label{fig.1}
\end{figure}
In Eq. 4, The  factor $\cos(\mathbf{{p}}\cdot \mathbf{{R}}/2)$
represents two-center interference\cite{Muth-Bohm}, and the term,
which is proportional to the internuclear distance $R$, leads to the
breakdown of translation invariance. In our calculations, this  term
is omitted according to Ref.\cite{cj,Milosevic,W. Becker}. Then, we
obtain
\begin{eqnarray}
\mathbf{d}^{mod}_{1s\sigma_{g}}(\mathbf{p})=N_{1s\sigma_{g}}[-2i\cos(\mathbf{{p}}\cdot
\mathbf{{R}}/2)\mathbf{{d}}_{1s}(\mathbf{{p}})].
\end{eqnarray}

In Fig. 1, we  plot the  harmonic spectra of 2D H$_2^+$ with $Z=1,
R=2$ a.u. at different laser intensities and orientation angles
$\theta$, obtained by the exact numerical calculation. The
comparison between the black and red curves in each subpannel of
Fig. 1 shows that the harmonic spectra have a broad region within
the HHG plateau with highly suppressed harmonic emission rate. The
center of the suppressed region shifts to higher harmonic order as
the orientation angle $\theta$ increases. The broad suppressed
regions arise from the  two-center interference effect in the
molecular HHG\cite{lein1,cyj}. But the locations of the minima in
the broad suppressed regions are difficult to identify in some
cases. For example, for $\theta=50^0$, in Fig. 1(c), the minimum is
clear at the 67th order.  Fig. 1(b) is not as clear as Fig. 1(c).

It has been revealed in Ref.\cite{cyj2} that the interference minima
in the molecular HHG spectra may shift as the laser intensity
changes. In particular, according to the simple point-emitter model
proposed in Ref\cite{lein2}, for H$_2^+$ with $R=2$ a.u. and
$\theta=50^0$, the predicted interference minimum is at the 51st
order. Compared to the observation of the 67th order in Fig. 1(c), a
large shift of about 16 orders occurs here. Based on the above
analyses and discussions, we expect that probing molecular structure
using the  pronounced interference-related minimum can not be
applicable in some cases.

\begin{figure}[t]
\begin{center}
\rotatebox{0}{\resizebox *{8.5cm}{8.cm} {\includegraphics
{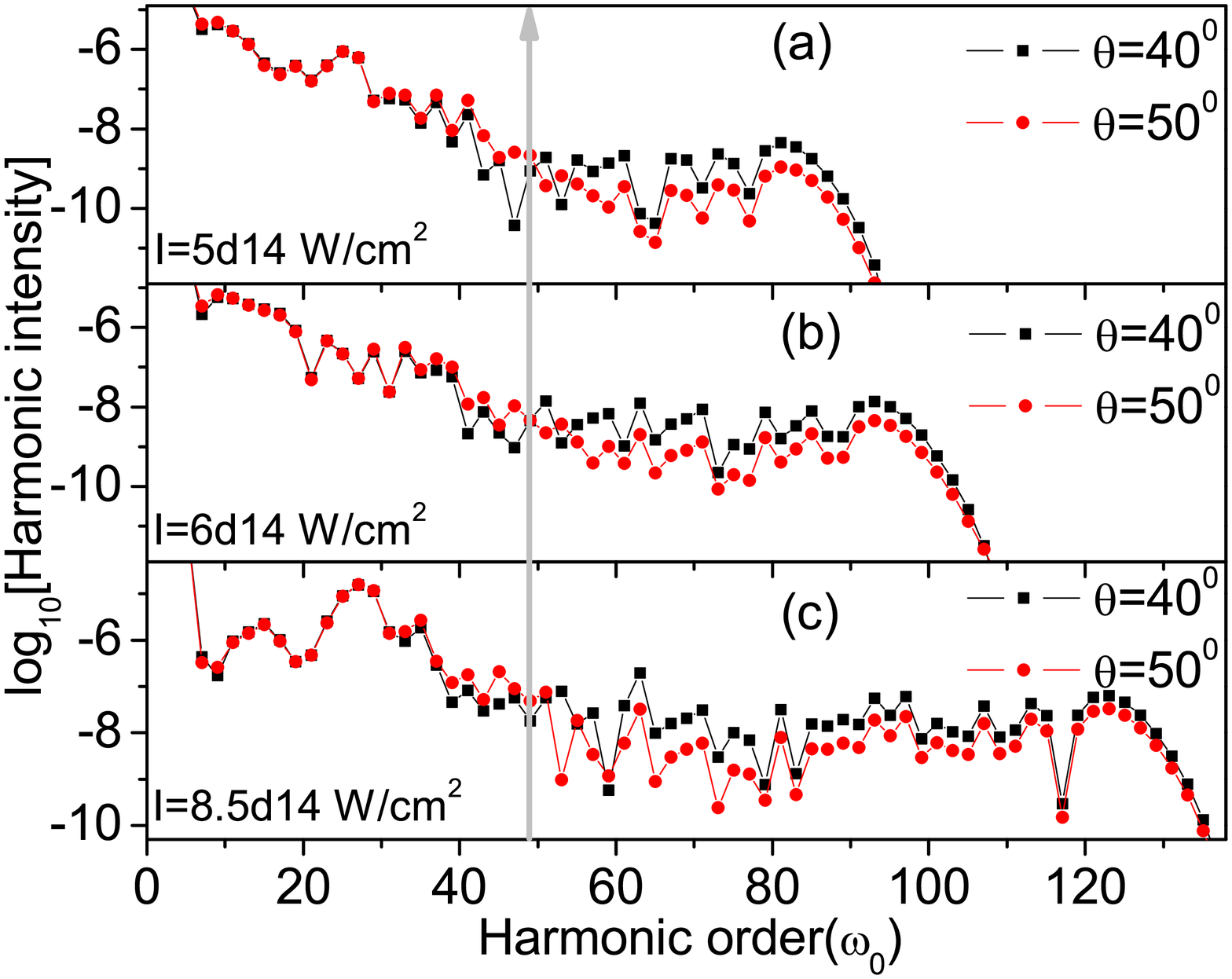}}}
\end{center}
\caption{Harmonic spectra  of 2D H$_2^+$  with $Z=1$, $R=2$ a.u. at
different laser intensities and orientation angles $\theta$,
calculated using the Lewenstein model.} \label{fig.2}
\end{figure}

However,  as we can see in each subpannel of Fig. 1, the two
harmonic spectra at different orientation angles $\theta$
demonstrate an intersection in the plateau region. For example, in
Fig. 1(c), the black curve for $\theta=40^0$ is lower from the 19th
to the 49th order, while the red curve for $\theta=50^0$ is lower
from the 49th to the 93rd order. The striking intersection of the
two curves is at the 49th order. As the laser intensity changes, the
intersection of the two curves is almost invariable, as indicated by
the vertical solid line. The calculated harmonic spectra at other
orientation angles $\theta$ also show the similar phenomena as those
revealed in Fig. 1. But the intersections of the harmonic curves
change as the orientation angles change. For example,  as presented
in Fig. 2, the intersection of two harmonic curves at $\theta=30^0$
and $\theta=50^0$ is at the 39th order (indicated by the vertical
solid line). This is different from that in Fig. 1.

Next, we concentrate on the physical mechanism behind these
phenomena. In Fig. 3, with the same parameters as in Fig. 1, we plot
the harmonic spectra calculated using the Lewenstein model Eq. 2 and
the modified transition dipole Eq. 5. One can see that the primary
characteristics of the harmonic spectra in Fig. 3 are analogous with
those of the corresponding curves in Fig. 1. For instance, in each
subpannel of Fig. 3, the intersection of  two harmonic spectra in
the plateau region is at the 49th order, as indicated by the
vertical solid curves. This is in agreement with that in Fig. 1. The
parallelism between the corresponding curves in Fig. 1 and Fig. 3
shows that the modified model, i.e., Eq. 2 with Eq. 5, is applicable
here for the description of the molecular HHG, especially for the
angle dependence of the HHG. This applicability is also consolidated
below.
\begin{figure}[t]
\begin{center}
\rotatebox{0}{\resizebox *{8.5cm}{8.cm} {\includegraphics
{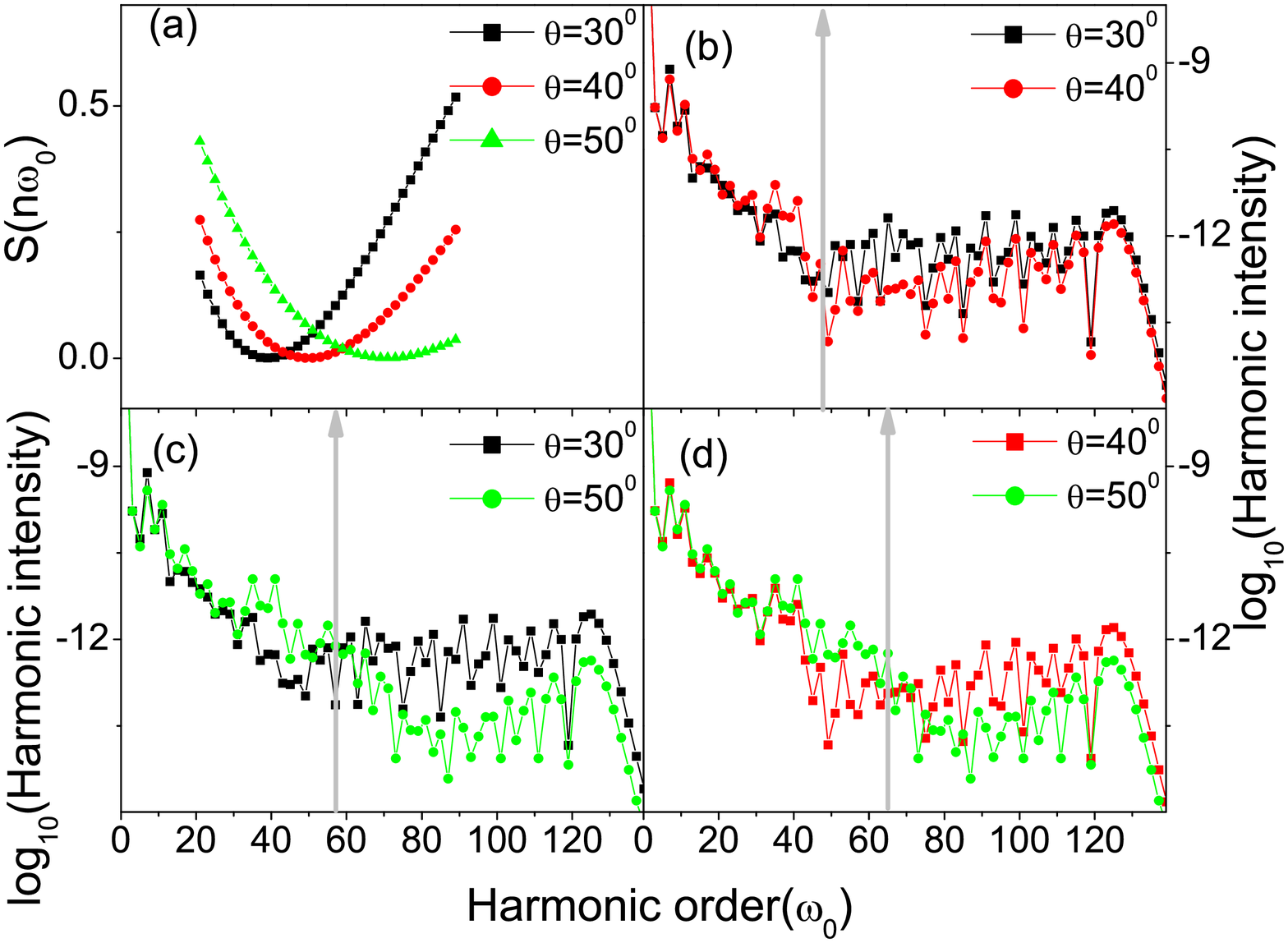}}}
\end{center}
\caption{Function curves of $S(n\omega_{0})$  and harmonic spectra
of 2D H$_2^+$  with $Z=1$, $R=1.7$ a.u. at different orientation
angles $\theta$.  (a): Function curves; (b), (c) and (d): harmonic
spectra obtained by the exact numerical calculation. The laser
intensity used here is $I=8.5\times10^{14}$W/cm$^{2}$.}
\label{fig.3}
\end{figure}

From the expressions of Eq. 2 and Eq. 5, we conjecture  that the
effects of two-center interference are responsible for the
intersections of the harmonic spectra at different orientation
angles. This conjecture is verified by the following analyses.
According to Ref.\cite{jitatani,Levesque}, the harmonic intensity
$F(\Omega=n\omega_0,\theta)$ along the laser polarization can be
written as $F(\Omega,\theta)\propto \Omega ^4 |
a(\Omega,\theta)D(\mathbf{p},\theta)|^2$, where $a(\Omega,\theta)$
is the spectral amplitude of the molecular recollision electronic
wave packet,
$D(\mathbf{p},\theta)=\vec{e}\cdot\mathbf{d}^{mod}_{1s\sigma_{g}}(\mathbf{p})$,
and $\Omega=E_\mathbf{p}+I_p$\cite{Corkum}. $E_\mathbf{p}$ is the
electronic kinetic energy. The spectral amplitude $a(\Omega,\theta)$
is closely related to the ionization process\cite{J.
Itatani,Lewenstein}. The transition dipole $D(\mathbf{p},\theta)$
corresponds to the recombination process. Both $a(\Omega,\theta)$
and $D(\mathbf{p},\theta)$  are alignment dependent\cite{cyj}. Here,
we focus on $D(\mathbf{p},\theta)$. The angle $\theta$ in the
expression of $D(\mathbf{p},\theta)$ is contained in the  $\cos$
function in Eq. 5. We extract the  $\cos$ function from
$D(\mathbf{p},\theta)$, and write its power  as
\begin{eqnarray}
S(n\omega_{0})=\cos(pR/2\cos\theta)\cos(pR/2\cos\theta),
\end{eqnarray}
with the dispersion relation $n\omega_{0}=p^2/2$, that considers the
electron acceleration before recombination\cite{jitatani}.  $n$ is
the harmonic order.

Based on the results shown in Fig. 3, we expect that Eq. 6 gives a
good description of the orientation dependence of the  HHG from
H$_2^+$. In Fig. 4 and Fig. 5, we show the comparisons between the
function curves of Eq. 6 and the corresponding harmonic spectra at
varied orientation angles $\theta$ and internuclear distances $R$.
These comparisons demonstrate the applicability of  Eq. 6 in the
prediction of the intersection of two harmonic spectra at different
orientation angles $\theta$.

\begin{figure}[t]
\begin{center}
\rotatebox{0}{\resizebox *{8.5cm}{8.cm} {\includegraphics
{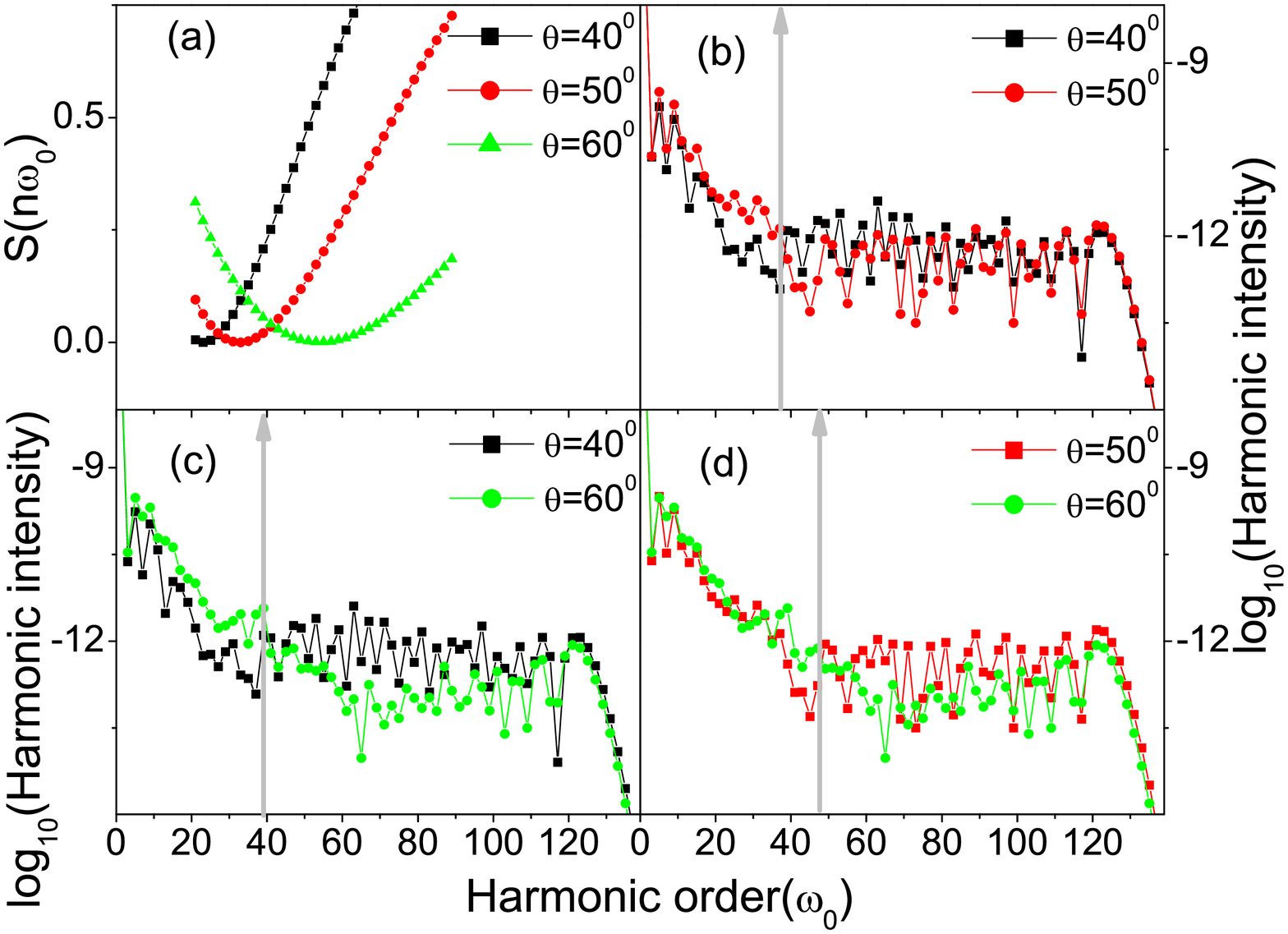}}}
\end{center}
\caption{Function curves of $S(n\omega_{0})$  and harmonic spectra
of 2D H$_2^+$  with $Z=1$, $R=2.5$ a.u. at different orientation
angles $\theta$.  (a): Function curves; (b), (c) and (d): harmonic
spectra obtained by the exact numerical calculation. The laser
intensity used here is $I=8.5\times10^{14}$W/cm$^{2}$.}
\label{fig.4}
\end{figure}

Specifically, for $R=1.7$ a.u., in Fig. 4(a), the formula predicts
that  the intersections of two harmonic spectra are at the 45th
order for $\theta=30^0$ and $\theta=40^0$, the 53rd order for
$\theta=30^0$ and $\theta=50^0$, and the 61st order for
$\theta=40^0$ and $\theta=50^0$. In the numerical cases in Fig.
4(b)-(d), the corresponding intersections are at the 47th order, the
57th order and the 65th order, respectively. For $R=2.5$ a.u., in
Fig. 5(a), the formula predicts that those are at the 29th order for
$\theta=40^0$ and $\theta=50^0$, the 35th order for $\theta=40^0$
and $\theta=60^0$, and the 43rd order for $\theta=50^0$ and
$\theta=60^0$. In Fig. 5(b)-(d), the corresponding intersections are
at the 37th order, the 39th order and the 47th order, respectively.
The large difference in the case of $\theta=40^0$ and $\theta=50^0$,
is due to the breakdown of the dispersion relation, used in our
simulation, in the low energy region\cite{Zhang,Xibin Zhou}. For
$R=2$ a.u., the formula also gives a good prediction.

In addition, one can see from  Fig. 4 and Fig. 5 that, the
interference-related minima are not distinct in the harmonic
spectra. It should be mentioned that our numerical method and
obtained HHG spectra are comparable to the previous work of
Ref.\cite{lein1,lein2,lein3}. However, in contrary to the claim of
Ref.\cite{lein1,lein2,lein3}, we find that the broad suppressed
regions in the harmonic spectra arise from the effects of two-center
interference in the HHG, while the locations of the
interference-related minima in the HHG spectra predicted in
Ref.\cite{lein1,lein2,lein3} are difficult to identify, as shown
above. Notice that in Ref.\cite{lein2,lein3}, a spectrum-smoothing
procedure is used to help the identification of the interference
minima. This procedure is not adopted in our analysis, since we
think it somehow  ambiguous and we expect a "direct" comparison of
the numerical observations to the experimental measurements.

There are two factors those could influence the position of the
interference minimum. First, two-center interference occurs not only
in the recombination process of the HHG, but also in the ionization
process of the HHG (into intermediate continuum states). The
interference factor $\cos(\mathbf{P}\cdot\mathbf{R}/2)$ is also
included in the transition dipole $\mathbf{d}_{ion}(\mathbf{p})$ in
Eq. 2 that corresponds to the ionization process. As a result, the
spectral amplitude $a(\Omega,\theta)$ of the molecular recollision
electronic wave packet, which is responsible for the fine structure
of the molecular HHG spectrum, may be affected by the
interference\cite{cyj}. Secondly, besides the ground state, the
first excited state can also contribute to the harmonic emission in
the broad suppressed region of the molecular HHG
spectrum\cite{cyj3}. Accordingly, the interference pattern may be
modulated by the population of the first excited state.

However, the parallelism between the predictions of Eq. 6 and the
numerical results, as discussed above, reveals that the
intersections of the HHG spectra could be less influenced by the two
factors. Particularly, compared to the interference-related minima,
the intersections are easier to identify. These suggest that  we can
measure the molecular bond length through the measurement of
harmonic spectra at different orientation angles $\theta$, using the
following equation
\begin{equation}
S(n_{s}\omega_{0},\theta_{1})=S(n_{s}\omega_{0},\theta_{2}),
\end{equation}
where $n_{s}$ is the harmonic order that corresponds to the
intersection of two harmonic spectra at the orientation angles
$\theta_1$ and $\theta_2$.

In conclusion, we have shown that due to the effects of two-center
interference in  the HHG, the harmonic spectra of H$_2^+$  at
different orientation angles demonstrate the striking intersections
in the high-frequency plateau region. The phenomena discussed here
are general. They are expected to appear in other species of
molecules. Our results can be useful for promising applications
allowing to use the "intersection" phenomenon as a prospective tool
to probe the molecular structure and dynamics.

This work is supported in part by Hong Kong Baptist University and
the Hong Kong Research Grants Council, and NNSF(No.10725521), 973
research program No.2006CB921400, 2007CB814800.


\begin{thebibliography}{2}
\bibitem [*] {} Liu$\_$Jie@iapcm.ac.cn

\bibitem{jitatani} J. Itatani, J. Levesque, D. Zeidler, Hiromichi Niikura, H.
Pepin, J. C. Kieffer, P. B. Corkum, and D. M. Villeneuve, Nature
\textbf{432}, 867(2004).

\bibitem{lein1} M. Lein, N. Hay, R. Velotta,  J. P. Marangos,  and P. L. Knight,  Phys. Rev. Lett \textbf{88},
183903, (2002).

\bibitem{lein2} M. Lein, N. Hay, R. Velotta,  J. P. Marangos,  and P. L. Knight,
Phys. Rev. A. \textbf{66}, 023805 (2002).

\bibitem{lein3} M. Lein, P. P. Corso, J.
P. Marangos, and P. L. Knight, Phys. Rev. A. \textbf{67}, 023819
(2003).

\bibitem{Bandrauk}G. Lagmago Kamta and A. D. Bandrauk, Phys. Rev. A \textbf{71}, 053407
(2005).
\bibitem{Tsuneto} T. Kanai, S. Minemoto, and H. Sakai, Nature \textbf{435},
470(2005).
\bibitem{vozzi}C. Vozzi, F. Calegari, E. Benedetti, J.-P. Caumes, G. Sansone, S. Stagira, M. Nisoli,
R. Torres, E. Heesel, N. Kajumba, J. P. Marangos, C. Altucci, and R.
Velotta, Phys. Rev. Lett. \textbf{95}, 153902 (2005).



\bibitem{J. Itatani} J. Itatani, D. Zeidler, J. Levesque, Michael Spanner, D. M. Villeneuve, and P. B. Corkum,
Phys. Rev. Lett \textbf{94}, 123902, (2005).
\bibitem{Levesque}J. Levesque, D. Zeidler, J. P. Marangos, P. B. Corkum, and
D. M. Villeneuve, Phys. Rev. Lett. \textbf{98}, 183903 (2007).

\bibitem{S. Patchkovskii}  S. Patchkovskii, Z. Zhao, T. Brabec, and D. M. Villeneuve,
Phys. Rev. Lett \textbf{97}, 123004, (2006).
\bibitem{R. Torres} R. Torres, N. Kajumba, Jonathan G. Underwood, J. S. Robinson, S.
Baker, J.W. G. Tisch, R. de Nalda, W. A. Bryan, R. Velotta, C.
Altucci, I. C. E. Turcu, and J. P. Marangos, Phys. Rev. Lett
\textbf{98}, 203007, (2007).
\bibitem{Le} Van-Hoang Le, Anh-Thu Le, Rui-Hua Xie, and C. D. Lin, Phys. Rev. A. \textbf{76},
013414 (2007)




\bibitem{Anh-Thu Le} Anh-Thu Le, X. M. Tong, and C. D. Lin,
Phys. Rev. A. \textbf{73} 041402(R) (2006).
\bibitem{Faria} C. Figueira de Morisson Faria, Phys. Rev. A. \textbf{76},
043407(2007).
\bibitem{Kanai}T. Kanai, S. Minemoto, and H. Sakai, Phys. Rev. Lett. \textbf{98}, 053002 (2007).
\bibitem{Ciappina} M. F. Ciappina, C. C. Chiril$\breve{a}$ and M. Lein, Phys. Rev. A. \textbf{75},
043405 (2007).
\bibitem{Wagner} N. Wagner {\it et al}., Phys. Rev. A. \textbf{76},
061403(R) (2007).
\bibitem{Ciappina2} M. F. Ciappina,  A. Becker and A. Jaro$\acute{n}$-Becker,
Phys. Rev. A. \textbf{76}, 063406 (2007).
\bibitem{Usachenko} V. I. Usachenko, P. E. Pyak, and S. I. Chu, Laser Phys. \textbf{16}, 1326 (2006).
\bibitem{cyj} Y. Chen, Y. Li, S. Yang, and J. Liu, Phys. Rev. A. \textbf{77}, 031402(R) (2008).



\bibitem{Lewenstein} M. Lewenstein, Ph. Balcou, M. Yu. Ivanov, Anne L'Huillier, and
P. B. Corkum, Phys. Rev. A \textbf{49}, 2117(1994).
\bibitem{cyj2} Y. J.
Chen and J. Liu, Phys. Rev. A. \textbf{77}, 013410 (2008).



\bibitem{Muth-Bohm} J. Muth-B\"{o}hm, A. Becker, and F. H. M. Faisal, Phys. Rev. Lett. \textbf{85}, 2280(4) (2000)

\bibitem{cj} J. Chen and S. G. Chen,  Phys. Rev. A. \textbf{75},
041402(R) (2007).
\bibitem{Milosevic} D. B. Milosevic,  Phys. Rev. A. \textbf{74}, 063404
(2006).
\bibitem{W. Becker} W. Becker, J. Chen, S. G. Chen, and  D. B. Milosevic,
Phys. Rev. A. \textbf{76}, 033403 (2007).



\bibitem{Corkum} P. B. Corkum, Phys. Rev.
Lett. \textbf{71}, 1994(1993).
\bibitem{Zhang} Zhangjin Chen, Toru Morishita, Anh-Thu Le, M. Wickenhauser, X. M. Tong, and C. D. Lin, Phys. Rev. A. \textbf{74} 053405 (2006).
\bibitem{Xibin Zhou}Xibin Zhou, Robynne Lock, Wen Li, Nick Wagner, Margaret M. Murnane, and Henry C. Kapteyn, Phys. Rev. Lett. \textbf{100}, 073902
(2008).
\bibitem{cyj3} Y. J. Chen,  J. Liu, and Bambi Hu,  High-order Harmonic Generation in a
Time-integrated Quantum Transition Picture, unpublished.





\end{thebibliography}
\end{document}